# Macroscale Molecular Communication in IoT-based Pipeline Inspection and Monitoring Applications: Preliminary Experiment and Mathematical Model

Muneer M. Al-Zubi, *Member, IEEE*, and Mohamed-Slim Alouini, *Fellow, IEEE*

*Abstract*—Today, pipeline networks serve as critical infrastructure for transporting materials such as water, gas, and oil. Modern technologies such as the Internet of Things (IoT), sensor nodes, and inspection robots enable efficient pipeline monitoring and inspection. They can help detect and monitor various conditions and defects in pipelines such as cracks, corrosion, leakage, pressure, flow, and temperature. Since most pipelines are buried underground, wireless communication links suffer from significant attenuation and noise due to harsh environmental conditions. In such systems, communication links are required between the sensor nodes as well as between the external control/monitoring unit or sensor node and the inspection robot inside the pipeline. In this paper, we propose a macroscale molecular communication (MC) system in the IoT-based pipeline inspection and monitoring networks to address this challenge. We develop a mathematical model and implement a preliminary experimental testbed to validate the system and demonstrate its feasibility by transmitting and reconstructing binary sequences using volatile organic compound (VOC) as an information signal. We examined the impact of various system parameters including airflow carrier velocity, released VOC velocity, emission duration, and bit duration. Results indicate that these parameters significantly influence the received molecular signal, emphasizing the need for optimal configuration. This work serves as a preliminary step for further research on the application of MC in IoT-based pipeline inspection and monitoring systems.

## I. INTRODUCTION

THE pipeline networks are widely used to transport various substances such as oil, gases, water, and more [1]. These networks require regular inspection, monitoring, and maintenance using advanced technologies like the Internet of Things (IoT), inspection robots, and wireless sensor nodes [2-6]. These technologies can detect and repair pipeline defects, including cracks, corrosion, and leaks, while monitoring fluid parameters like quality, consumption, pressure, flow, and temperature [7-11]. Today, IoT equipment for pipeline monitoring applications are available in the market, such as the cellular IoT gateway manufactured by Beilai Technology [12]. IoT-based pipeline monitoring systems typically use sensor nodes positioned along the pipeline to measure parameters like pressure, flow rate, and temperature.

The sensor nodes then transmit the collected data to the gateway (data acquisition device) via wireless links. The gateway processes and forwards the sensor data to a base station (BS). The BS then transmits the data to the control and monitoring center via the Internet. Additionally, inspection robots can perform pipeline inspection and maintenance tasks, which require communication links with sensor nodes or external control/monitoring units. Wireless communication technologies used for IoT-based monitoring and inspection of aboveground pipelines typically encounter minimal challenges related to radio wave propagation and attenuation. However, most pipelines are buried underground (as illustrated in Fig. 1), where wireless links face severe signal attenuation and noise due to the harsh underground environment. This environment includes soil, rocks, moisture, metal pipes, and other obstructive materials [13-16]. This signal degradation impacts three critical wireless communication links: (1) the connection between the external gateway and underground sensor nodes, (2) the link between the external control unit and the inspection robot, and (3) the inter-node links in mesh-topology configurations. Alternative communication techniques, such as optical and wired communication, have also been proposed for in-pipe inspection robots [17, 18]. However, these methods present limitations: optical communication requires line-of-sight (LOS) conditions, while wired solutions depend on physical cable connections.

Inspired by nature, molecular communication (MC) is a promising and emerging solution for data communication in pipeline environments where conventional communication methods are inefficient or impractical [19, 20]. In recent years, MC has received increasing interest among researchers worldwide. In MC systems, data is encoded and transmitted from the transmitter (TX) to the receiver (RX) using chemical molecules that act as information carriers [21]. In MC, chemical molecules transport via diffusion and advection over both nanoscale (i.e., nm-μm) and macroscale (i.e., cm-meters) distances. Macroscale MC emerges as a promising solution for industrial and environmental applications, including environmental monitoring, oil/gas leak detection, and in-pipeline data communication [22]. Therefore, macroscale MC can offer a promising solution for addressing communication

The authors are with the Computer, Electrical and Mathematical Science and Engineering Division (CEMSE), King Abdullah University of Science and Technology (KAUST), Thuwal 23955-6900, Makkah Province, Saudi Arabia (e-mail: muneer.zubi@kaust.edu.sa; slim.alouini@kaust.edu.sa).

challenges inside the pipelines. This enables the following data communication links: communication between in-pipe sensor nodes, communication between in-pipe inspection robots and external MC control/monitoring units, and communication between in-pipe inspection robots and in-pipe sensor nodes as shown in Fig. 1. Most studies in the literature focus on nano-to micro-scale MC [19, 23-25], while only a few works explore MC over macro-scale ranges [26-31]. Macroscale MC is relatively a new research area that requires distinct considerations in designing and modelling the system components including the TX, the channel, and the RX. A macro-scale MC platform is implemented by releasing alcohol as signaling molecules in open air which is detected using a metal-oxide (MOX) sensor at the RX [31]. Few studies have explored macro-scale MC systems utilizing advection-diffusion channels, including both open-air environments and enclosed pipes [26, 27, 32, 33]. These implementations employ an odour generator as TX and a mass spectrometer as RX for message detection. In other studies, macroscale MC systems are implemented in water-based environments using hydrogen ions as information carriers that are detected by pH sensors [34, 35]. In [28], the authors developed an MC testbed that employs magnetic nanoparticles as information carriers, transported through microfluidic channels via water flow and detected using a susceptometer. Additionally, researchers have investigated optical-based macro-scale MC systems operating in both laminar and turbulent flow regimes, employing planar laser-induced fluorescence (PLIF) and particle image velocimetry (PIV) as detection techniques [36-38]. In [39], the MC system was implemented using a chemical vapor emitter as TX and photoionization detectors (PIDs) as RX, with signal propagation through a turbulent flow tube channel. While the aforementioned studies make significant contributions to macro-scale MC research, they have not specifically explored its application in IoT-based pipeline inspection and monitoring networks. These works do not address MC links between in-pipe sensor nodes or between inspection robots and external MC control/monitoring units (or other sensor nodes), as illustrated in Fig. 1. Furthermore, accurate modelling of macro-scale MC systems must simultaneously account for dynamic channel conditions, as these critically influence both the carrier flow profile and the spatiotemporal distribution of information molecules. These critical conditions encompass turbulent flow regimes, pipe boundary conditions, pipeline topology, and emission process characteristics that determine the system's communication performance. This modelling challenge can be addressed using numerical solutions via computational fluid dynamics (CFD) software packages such as ANSYS Fluent and COMSOL Multiphysics.

In this paper, we propose a macroscale MC system in an IoT-based pipeline inspection and monitoring network for providing communication links between sensor nodes as well as between an inspection robot and a external control/monitoring unit or sensor node. We have developed a mathematical model and experimental testbed for the proposed system inside an L-shaped pipe as a preliminary demonstration and proof of concept to verify the developed model and investigate the feasibility of the proposed system. Although the current implementation achieves a low data rate (on the order of bits), this work demonstrates that future optimizations can enhance system performance. Through comprehensive modelling and experimental validation, this paper establishes a foundation and starting point for researchers to further advance the proposed system. The main contributions of this paper include the following:

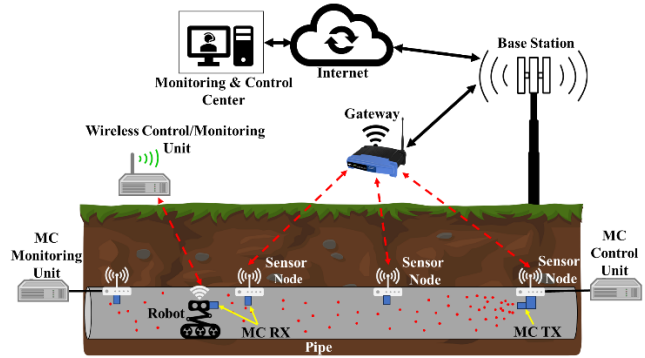

**Fig. 1**. The architecture of IoT-based pipeline inspection and monitoring system using sensor nodes and inspection robots via MC and wireless communication.

(1) We propose a macroscale MC system that utilizes volatile organic compound (VOC) gas as an information signal, designed for integration into IoT-based pipeline inspection and monitoring systems. As illustrated in Fig. 1, the proposed MC system enables three key communication scenarios: (i) between in-pipe sensor nodes, (ii) between an external control/monitoring unit and an in-pipe robot, and (iii) between an in-pipe sensor node and an inspection robot. This system can transmit short messages or control commands (e.g., task instructions for robots) and facilitate simple data exchange between sensor nodes. By complementing existing wireless communication systems, this MC system addresses critical connectivity challenges within pipelines.

(2) We develop a mathematical model based on the CFD approach for the proposed system which consists of three main components that can be abstracted as TX, channel, and RX as shown in Fig. 2(a). In this system, the TX and RX may represent either a sensor node communicating with another node or an external control/monitoring unit (or an in-pipe sensor node) communicating with an inspection robot inside the pipe. As shown in Fig. 2(a), the TX controls the release of VOC (i.e., information signal) into the main pipe using a flow/spray control unit connected to the branch pipe. The RX consists of a VOC sensor integrated with a signal processing unit, mounted either on an inspection robot or a receiving sensor node. The channel corresponds to the main pipeline, which propagates the information signal from the TX to the RX via bulk airflow (the carrier signal).

(3) The governing equations in the mathematical model are solved numerically through transient analysis, incorporating coupled solutions of the turbulence model, transport model, and conservation equations. To accurately represent real-world conditions, we model the channel's irregular flow as turbulent flow as evidenced by the large values of the Reynolds number. This yields more accurate results compared to steady-state analyses and decoupled solution methods.

(4) We developed an experimental testbed using commercial off-the-shelf (COTS) components to validate the proposed system and evaluate its feasibility, as illustrated in Fig. 2(b). The testbed employs VOC taken from an air freshener as an information carrier. A sequence of binary bits is transmitted by releasing a modulated information signal in the pipe channel and reconstructed successfully at the RX. System performance is evaluated by transmitting various binary sequences and analyzing the bit error rate (BER).

(5) We have validated the mathematical model against experimental measurements obtained from the testbed. Furthermore, we conducted a comprehensive analysis of key system parameters affecting the received molecular signal, including the emission pulse duration, the VOC velocity (i.e., the information signal), the airflow velocity (i.e., the carrier signal), and the bit duration.

The remainder of this paper is organized as follows. Section II presents an overview of the proposed system architecture. Section III details the mathematical modelling framework, including formulation and numerical analysis. Section IV describes experimental testbed implementation. Section V presents and discusses the key results. Finally, the paper is concluded in Section VI.

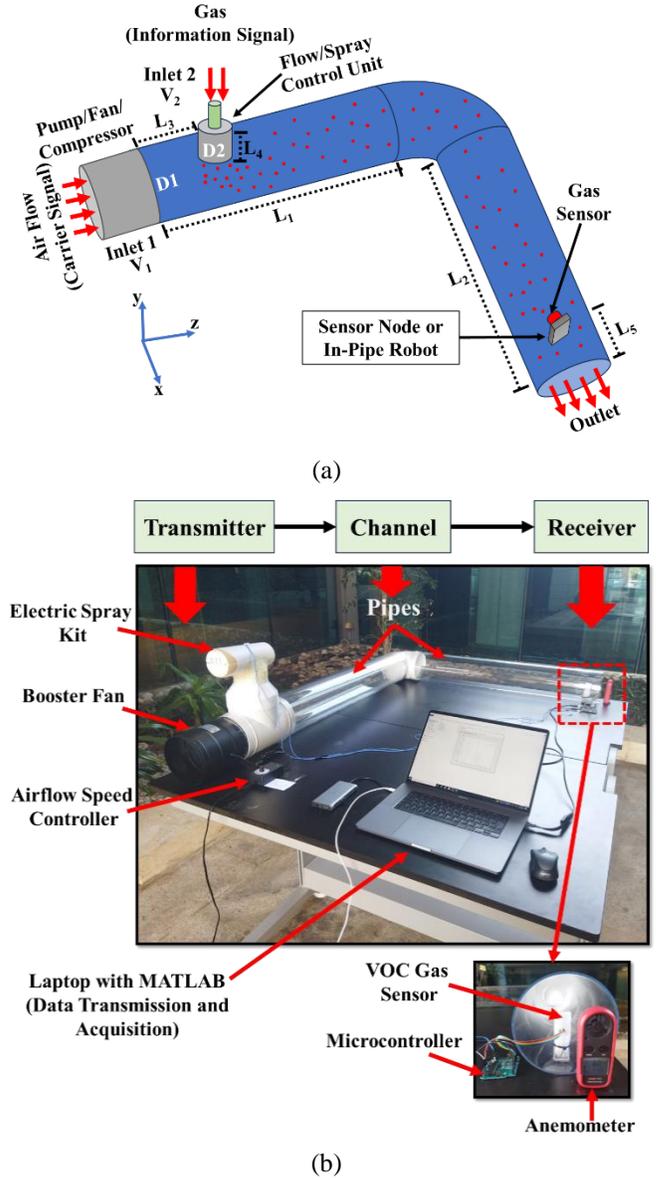

(a)

(b)

**Fig. 2.** A graphical illustration of (a) the architecture and geometry of the proposed MC system and (b) the experimental testbed setup.

## TABLE I
### THE PIPE DIMENSION PARAMETERS USED IN THIS SYSTEM.

| Parameter | Value (cm) | Description |
|---|---|---|
| $D_1$ | 15 | Diameter of the main pipe. |
| $D_2$ | 15 | Diameter of the branch pipe. |
| $L_1$ | 115 | Length of segment 1 of the main pipe. |
| $L_2$ | 91 | Length of segment 2 of the main pipe. |
| $L_3$ | 10 | Distance between the inlet-1 and the branch pipe. |
| $L_4$ | 13 | Length of the branch pipe. |
| $L_5$ | 7 | Distance between the robot or sensor node and the outlet. |

## II. SYSTEM ARCHITECTURE OVERVIEW

This section describes the structure and components of the proposed MC system. This system provides an alternative communication solution that enables communication between in-pipe sensor nodes as well as between the in-pipe inspection robot and either a remote/monitoring unit or sensor node. In this system, the pipeline is assumed to be empty of any materials or substances during the maintenance process. The robot is assumed to be stationary during communication to analyze the received molecular signal. However, the impact of robot mobility on the received molecular signal is left for future work. As shown in Fig. 2(a), an L-shaped pipeline segment, a common configuration in real-world pipeline networks, is chosen to examine the proposed MC system. Nevertheless, the same framework can be extended to other pipeline geometries.

The pipe topology consists of a small branch pipe connected to the main pipe that is used to transport fluid substances. The pipe dimension parameters used in this system are listed in Table I. In this system, the TX acts as a sensor node or an MC control unit, as illustrated in Fig. 1. The data message or command is emitted as a bitstream by the TX via releasing the information signal (i.e., the VOC pulses). The TX releases VOC pulses through a small branch pipe connected to a spray control unit, which controls flow velocity and releasing duration of VOC gas during specified time intervals. In this system, bit-1 transmission is achieved by emitting a VOC pulse at a velocity $V_2$ for a short emission duration $T_e$, while bit-0 transmission corresponds to stopping the VOC release. This can be seen as an on-off keying (OOK) modulation which is classified under the concentration-based modulation techniques. It is worth mentioning here that other modulation methods can be used flexibly in this system such as type-based, timing-based, and concentration-based modulation techniques. The emitted information signal propagates through the channel (main pipe) via the carrier signal (continuous airflow) to be finally detected by the VOC sensor at the RX. Here, the RX represents another sensor node or an in-pipe inspection robot. The gas sensor measures the VOC concentration, which is subsequently demodulated and processed into binary data output. A carrier signal (airflow) generated at pipe inlet-1 by an air pump, compressor, or booster fan is used to speed up the MC system by reducing the arrival time of the information signal at the RX and consequently increasing the achievable data rate. This system has been theoretically modelled and analyzed using the CFD approach, as will be demonstrated in the next section.

To validate the developed model and evaluate the system feasibility, we implemented an experimental testbed using COTS components, as illustrated in Fig. 2(b). In this setup, VOC gas taken from air freshener spray serves as the information carrier to transmit messages/commands (i.e., bit sequences) to the RX through the pipeline channel. It is worth mentioning that other chemical gases can be used as information signals in the proposed system with minimal modification to the system structure. The experimental testbed will be discussed in detail in Section IV.

## III. MATHEMATICAL MODEL

In this section, we present the mathematical formulation and numerical analysis of the proposed MC system described earlier. The process of releasing the information signal (i.e., the VOC pulses) into the channel (i.e., the main pipe) with the carrier signal (i.e., the airflow) can be seen as a mixing of VOC and air. However, since the VOC concentration entering the main pipe is significantly lower than that of air, we approximate this process as the transportation of VOC gas in airflow, governed by advection and diffusion. In this system, heat transfer is neglected due to minimal temperature variations. Also, incompressible flow is assumed with constant fluid density, as the Mach number (M) satisfies M < 0.1.

### A. The governing equations

The system is governed by the following equations: the momentum equation, the continuity equation, the k-$\varepsilon$ turbulence model, and the species transport equation. The momentum and continuity equations are derived from the Navier-Stokes equations. The general form of the continuity (mass conservation) equation is given as [40],

$$\frac{\partial \rho}{\partial t} + \nabla \cdot (\rho \vec{u}) = 0 \qquad (1)$$

where $\rho$ is the fluid (gas mixture) density in (kg/m$^3$), $\vec{u}$ is the velocity vector along (x, y, z) directions in (m/s), $t$ is the flow time in (s), and $\nabla$ is the del (gradient) operator. The momentum equation is given as [40],

$$\frac{\partial}{\partial t}(\rho \vec{u}) + \nabla \cdot (\rho \vec{u} \vec{u}) = -\nabla p + \nabla \cdot (\bar{\bar{\tau}}) + \rho \vec{g} + \vec{F} \qquad (2)$$

where $p$ is the fluid static pressure in (Pa), $\bar{\bar{\tau}}$ is the stress tensor,

$\vec{g}$ is gravitational acceleration in (m/s$^2$), $\rho \vec{g}$ is the gravitational body force acting on the gas flow, and $\vec{F}$ is the external body force. The stress tensor is expressed as:

$$\bar{\bar{\tau}} = \mu \left[ (\nabla \vec{u} + \nabla \vec{u}^T) - \frac{2}{3} \nabla \cdot \vec{u} I \right] \qquad (3)$$

where $\mu$ is the molecular viscosity in (Pa.s), $I$ is the unit tensor, and the second term on the right-hand side is the effect of volume dilation.

The transport of the information signal (VOC gas) under the influence of the carrier signal (airflow) in the main pipe can be modelled using the following species transport (advection-diffusion) equation [40]:

$$\frac{\partial}{\partial t}(\rho Y) + \nabla \cdot (\rho \vec{u} Y) = \nabla \cdot (\rho D \nabla Y) + S \qquad (4)$$

where $Y$ is the mass fraction of the molecular signal at location (x, y, z) at time $t$, $D$ is the diffusion coefficient of VOC in the air in (m$^2$/s), and $S$ is the source term.

In this study, we used Glade® air freshener as the VOC source, which contains acetone as a major ingredient that is used as a carrier to help dissolve fragrance to dispense into the air. Therefore, we use the acetone characteristic to model the VOC transport in this model. In this macro-scale system, the dominant force that transports the information signal in the pipe is the airflow (advection) rather than the diffusion where the Peclet number Pe>>1.

Turbulent flow is described by the rapid fluctuation of velocity fields in both space and time. The gas flow in this system is considered a turbulent flow which can be observed from the high Reynolds number (Re), i.e., an inlet velocity (V$_1$) between 1-5 m/s results in $Re$ values between 10,680 and 53,415.

$$Re = \frac{\rho \cdot u \cdot D_1}{\mu} \qquad (5)$$

where D$_1$ is the main pipe diameter in meters.

In this work, we use the standard k-$\varepsilon$ turbulence model built into the ANSYS Fluent to model the turbulent flow conditions. The standard $k$-$\varepsilon$ turbulence model has become the most powerful and common model used in CFD analysis for turbulent flow calculations in practical and industrial engineering applications [41, 42]. This model was experimentally found to be an accurate model for high-velocity conditions. It is a semi-empirical model that provides a general description of turbulence by solving two transport equations for kinetic energy ($k$) and its dissipation rate ($\varepsilon$) as follows:

$$\frac{\partial}{\partial t}(\rho k) + \frac{\partial}{\partial x_i}(\rho k u_i) = \frac{\partial}{\partial x_i}\left[\left(\mu + \frac{\mu_t}{\sigma_k}\right)\frac{\partial k}{\partial x_i}\right] \qquad (6)$$
$$+ G_k + G_b - \rho \varepsilon - Y_m + S_k$$

$$\frac{\partial}{\partial t}(\rho \varepsilon) + \frac{\partial}{\partial x_i}(\rho \varepsilon u_i) = \frac{\partial}{\partial x_i}\left[\left(\mu + \frac{\mu_t}{\sigma_\varepsilon}\right)\frac{\partial \varepsilon}{\partial x_i}\right] \qquad (7)$$
$$+ C_{1\varepsilon}\frac{\varepsilon}{k}(G_k + C_{3\varepsilon}G_b) - C_{2\varepsilon}\rho\frac{\varepsilon^2}{k} + S_\varepsilon$$

where the terms ($G_k$, $G_b$) are the generation of turbulence kinetic energy due to the mean velocity gradients and buoyancy, respectively, and $Y_m$ is the contribution of the fluctuating dilatation in compressible turbulence to the overall dissipation rate. The terms $S_k$ and $S_\epsilon$ are user-defined source terms. The factors $\sigma_k$ and $\sigma_\epsilon$ are the turbulent Prandtl numbers for $k$ and $\epsilon$, respectively. The turbulent (eddy) viscosity ($\mu_t$) is expressed as

$$\mu_t = \rho C_\mu k^2 \epsilon^{-1} \qquad (8)$$

where the default values of the constants $C_{1\epsilon}$, $C_{2\epsilon}$, $C_{3\epsilon}$, $C_\mu$, $\sigma_k$, and $\sigma_\epsilon$ have been determined from experiments of a wide range of turbulent flows and are equal to 1.44, 1.92, 0, 0.09, 1, and 1.3, respectively. This turbulence model is numerically solved in ANSYS Fluent, and the solution methodology is out of the scope of this work.

### B. Initial and boundary conditions

The initial and boundary conditions are required to describe the system behaviour on the domain boundaries and the system state at the initial time. The initial and boundary conditions are defined at the different parts of the system shown in Fig. 2(a) as follows:

**Inlet 1**: the carrier signal (airflow) enters the main pipe continuously from this inlet. Therefore, velocity inlet boundary condition with a fixed velocity profile normal to the boundary is applied as follows:

$$(\vec{u}_x, \vec{u}_y, \vec{u}_z) = (0, 0, V_1) \text{ m/s} \qquad (9)$$

where $V_1$ is the airflow velocity in the main pipe in (m/s). In addition, the atmospheric pressure is chosen as an initial condition at this inlet as

$$p(t_0) = 101,325 \text{ Pa} \qquad (10)$$

The inlet turbulent intensity is chosen as 5% and the hydraulic diameter is equal to the main pipe diameter ($D_1$).

**Inlet 2**: the information signal (VOC gas) enters the main pipe from this inlet via the branch pipe. Because the VOC signal is emitted as pulses (0 and 1 bits), a velocity inlet boundary condition using a rectangular pulse velocity profile is applied as follows:

$$\vec{u}_y = \begin{cases} V_2, & nT_b < t \le (nT_b + T_e) \\ 0, & otherwise \end{cases} \text{ m/s} \qquad (11)$$

$$(\vec{u}_x, \vec{u}_z) = (0, 0) \text{ m/s}$$

where $V_2$ is the flow velocity of VOC gas in the branch pipe in (m/s), $T_b$ is the bit duration in (s), $T_e$ is the emission duration of the VOC signal in (s), $n = 0, 1, 2,...,N$ is the bit index and N is the number of the transmitted bits. The general form of mass and molar flow rates of VOC at inlet 2 can be obtained from the velocity profile as follows.

$$\dot{m} = \begin{cases} \rho_{VOC} \cdot A \cdot V_2, & nT_b < t \le (nT_b + T_e) \\ 0, & otherwise \end{cases} \qquad (12)$$

$$\dot{n} = \begin{cases} \dot{m}/MW_{VOC}, & nT_b < t \le (nT_b + T_e) \\ 0, & otherwise \end{cases} \qquad (13)$$

where $\dot{m}$ is the mass flow rate in (g/s), $\dot{n}$ is the molar flow rate in (mol/s), A is the cross-sectional area of the branch pipe in (m²), $MW_{VOC}$ is the molar weight of VOC gas in (g/mol), and $\rho_{VOC}$ is the VOC density in (kg/m³). Similar to inlet 1, the atmospheric pressure is chosen as an initial condition at this inlet as given in Eq. (10). Moreover, the inlet turbulent intensity is chosen as 5% and the hydraulic diameter is equal to the branch pipe diameter ($D_2$).

**Outlet:** the gas mixture (air and VOC) leaves the main pipe from the outlet which is open to the atmosphere. Thus, zero pressure boundary condition is used at the outlet (i.e., zero-gauge pressure represents the atmospheric pressure that is equal to 101,325 Pa).

**Walls:** nonslip boundary condition is defined at the pipe wall boundaries. Thus, the velocity components at the pipe wall are set to zero.

$$\vec{u}(x, y, z, t) = (0, 0, 0) \text{ m/s} \qquad (14)$$

### C. Mesh generation and model geometry

In this work, a 3-D cylindrical L-shaped main pipe with an internal diameter ($D_1$) is connected to a branch pipe of diameter ($D_2$), as shown in Fig. 2(a). The model geometry is created using ANSYS DesignModeler. Since mesh size can significantly influence computational efficiency and the accuracy of the results, a balance must be achieved between resolution and processing time. Generally, a finer mesh yields higher accuracy but requires longer computational time. Therefore, a mesh independence study, aka., a mesh sensitivity study, was conducted to evaluate the effect of different grid resolutions on the transient solution of both the VOC concentration and the airflow velocity profile, ensuring reliable results. Based on this analysis, a mesh consisting of 110,160 elements with an average size of 15 mm was selected to build the geometry. An unstructured tetrahedral mesh type was applied across the domain. Moreover, we refined the mesh at the inlet and outlet faces as well as at the pipe wall to resolve boundary layer separation and to handle the rapid change in the flow velocity and flow mass rate. Near the pipe wall, the mesh was further refined using inflation with five inflation layers.

### D. Numerical analysis method and setup

The available methods for modeling MC systems include stochastic simulation, analytical modelling, and numerical analysis via the finite element/volume method. The mathematical model describing this system is complex and impossible to be solved analytically. Moreover, macroscale stochastic particle-based simulation in such a system is impractical due to large geometry and complex flow conditions (e.g., turbulent flow and boundary effects) that require complex stochastic modelling and heavy computation with long processing time. In contrast, numerical simulation using CFD software like ANSYS Fluent provides a more feasible approach for such complex MC systems. Therefore, in this study,

ANSYS Fluent (2022 R2) is employed to numerically solve the governing equations and analyze the system's behaviour.

The solution is obtained by simultaneously solving the species transport model and the turbulent flow model. ANSYS Fluent uses the finite volume method (FVM) to solve the governing partial differential equations (PDEs). In the solver settings, a pressure-based transient solver with double precision is selected to ensure accuracy. For pressure-velocity coupling, the coupled scheme is applied. In spatial discretization, the least squares cell-based approach is used to obtain the gradients. Also, a second-order upwind scheme is used for momentum, turbulent kinetic energy, turbulent dissipation rate, and VOC gas concentration. These settings collectively ensure a robust and accurate simulation of the complex flow and transport phenomena in the system.

The numerical simulation employs a strict convergence criterion of $10^{-6}$ for residual monitoring. The boundary and initial conditions, detailed in subsection B, are applied in ANSYS Fluent. To model the time-varying VOC emission process, a user-defined function is used, which contains the pulsed emission of binary data bits. For turbulence modelling, the standard k-ε turbulence model is selected with standard wall functions for near-wall treatment. The species transport model utilizes a VOC-air mixture approach with the volume-weighted mixing law, enabling ANSYS Fluent to automatically compute the gas mixture density throughout the simulation. Finally, the VOC concentration is continuously monitored in the sensing domain at the RX location (i.e., the gas sensor on a robot or a sensor node), see Fig. 1 and Fig. 2(a). The system parameters used in the CFD-based numerical analysis in ANSYS Fluent are listed in Table II.

## IV. EXPERIMENTAL TESTBED

An experimental testbed is implemented to verify the developed theoretical model and examine the feasibility of the proposed system as shown in Fig. 2(b). This system has the flexibility to change the gas used as an information signal and to choose different pipe dimensions. We use the VOC gas found in the air freshener as an information signal in the experimental testbed. To prevent VOC accumulation in the laboratory, the experiment was conducted outdoors. The testbed can be divided into three main parts as follows:

**Transmitter:** In the proposed IoT-based pipeline monitoring and inspection system, the transmitter acts as an MC sensor node or as an external control/monitoring unit. In this demonstration experiment, the transmitter consists of an electric air freshener spray kit which is modified to be controlled via an external microcontroller (Arduino UNO). The electric sprayer is connected to the main pipe through a branch pipe as illustrated in Fig. 2(b). The VOC contents in Glade® air freshener is used as an information signal. The size of the Glade® air freshener bottle is 269ml which provides up to 2400 sprays. The air fresheners contain the following chemicals: acetone, isobutane, propane, and fragrance [43]. In Glade® air freshener, acetone represents the major ingredient used as a carrier that helps dissolve a fragrance, so it dispenses into the air. The transmitted binary bits are generated using a MATLAB code that runs on a laptop connected to the microcontroller. The microcontroller controls the opening and closing of the electric sprayer to release the VOC via OOK modulation depending on the generated bits. For instance, the sprayer opens for a duration ($T_e$) for sending bit "1" while it remains closed for sending bit "0". The used electric sprayer in this experiment has a fixed emission duration which is approximately equal to one second ($T_e$=1s). However, it is possible to replace this sprayer with a more advanced unit that provides control capability for the release duration.

TABLE II
THE SYSTEM PARAMETERS USED IN THE NUMERICAL ANALYSIS IN ANSYS FLUENT.

| Parameter | Value | Unit | Description |
|---|---|---|---|
| $\rho_{air}$ | 1.225 | kg/m³ | Air density |
| $\rho_{voc}$ | 2 | kg/m³ | VOC density |
| $\mu_{air}$ | 1.7894×10⁻⁵ | kg/(m.s) | Air dynamic viscosity |
| $\mu_{voc}$ | 33.1×10⁻⁵ | kg/(m.s) | VOC dynamic viscosity |
| $\mu_{voc-air}$ | 1.72×10⁻⁵ | kg/(m.s) | Dynamic viscosity of Air/VOC mixture |
| $D_{voc-air}$ | 1.2×10⁻⁵ | m²/s | Diffusion coefficient of VOC in air |
| $V_1$ | 1-6 | m/s | Airflow velocity at inlet 1 |
| $V_2$ | 1-5 | m/s | VOC flow velocity at inlet 2 |
| $T_e$ | 0.5-2 | s | Emission duration of VOC for sending bit 1 |
| $T_b$ | 1-5 | s | Bit duration |
| $MW_{voc}$ | 58.08 | g/mol | The molecular weight of VOC |
| $MW_{air}$ | 28.96 | g/mol | The molecular weight of air |
| $p$ | 101,325 | Pa | Operation pressure |

**Channel:** The channel in this system consists of the pipe through which the VOC signal propagates from TX to the VOC sensor at RX. This pipe is made of transparent polycarbonate material. The pipe dimensions used in this testbed are given in Table I. Also, the carrier signal (bulk airflow) can be considered as a part of the channel which speeds up the information signal (VOC) flow in the pipe to increase the bit rate. The bulk airflow is generated using a booster fan with a speed controller (Hon&Guan brand). We measured the generated airflow velocity at the pipe outlet using a calibrated anemometer (Wintact WT816), obtaining values ranging from 3.0 m/s to 6.4 m/s depending on the fan speed setting.

**Receiver:** In the proposed system, the receiver acts as an MC sensor node or in-pipe robot. In this testbed, the receiver implementation comprises a VOC gas sensor interfaced with an Arduino UNO microcontroller. As illustrated in Fig. 2(b), the gas sensor is positioned near the pipe outlet. We have tested multiple metal-oxide (MOX) gas sensors, including MQ-138 and BME680, to check their sensitivity and response/recovery time in detecting the released VOCs. The experimental results show that the BME680 sensor provides the best sensitivity and response to the VOC signal with response time < 1s. The BME680 sensor is the first gas sensor that integrates high-linearity and high-accuracy gas, pressure, humidity, and temperature sensors. This sensor can detect a broad range of gases and VOCs such as acetone, ethanol, and carbon monoxide. We use the CJMCU680 board module that integrates the BME680 sensor which supports SPI and I²C communication interfaces with the microcontroller. The BME680 sensor provides a raw output corresponding to the resistance $R$ of its heated MOX element in (k$\Omega$). This resistance exhibits an inverse proportionality to VOC concentration which decreases as VOC levels increase, and vice versa. The raw resistance data acquired from the gas sensor is transferred to a laptop for processing. We then compute the corresponding conductance $G$ in k$\Omega^{-1}$ through the inverse relationship G = 1/R and visualize the results using MATLAB. The conductance is directly proportional to the VOC concentration in the air.

## V. RESULTS AND DISCUSSION

In this section, we present and analyze the numerical results obtained from solving the mathematical model of the proposed system. These results are verified and validated against experimental measurements from the testbed. Understanding how key parameters influence flow behaviour and the molecular received signal at the RX is critical for the macroscale MC system. Thus, we examine the effect of the various system parameters on the molecular received signal at the RX including the emission duration of the information signal, the bit duration, the information signal velocity, and the carrier signal velocity. We refer to the information signal concentration, detected by the gas sensor, as a molecular received signal.

In the experimental testbed, we normalized the VOC gas sensor response by converting raw resistance measurements (in ohms) to conductance values scaled to their maximum. Conductance is directly proportional to the VOC concentration in the air. The system has been modelled in ANSYS Fluent (2022 R2) software. In this system, the information signal propagates through the pipe primarily via airflow transport rather than diffusion, as indicated by the high Péclet number (Pe>>1).

For direct comparison with experimental results, the numerically obtained molecular received signal was normalized to its maximum value. Figure 3(a) presents the temporal normalized sensor response for different airflow velocities (V₁) following transmission of bit "1". The results demonstrate excellent agreement between numerical and experimental data during the response period including peak amplitude and peak time. The gas sensor exhibits rapid response characteristics, with a response time of under one second. It is worth mentioning that the peak time and peak amplitude are two critical parameters for accurate bit reconstruction. However, experimental results show a longer decay phase compared to simulations, attributable to the sensor's recovery time required to return to baseline. Also, we can observe shorter peak time and signal width as the airflow velocity increases because the gas will reach and leave the sensor faster. As expected, there is a channel delay for the emitted information signal to reach the gas sensor at the RX which decreases as the airflow speed increases from 3m/s to 6m/s.

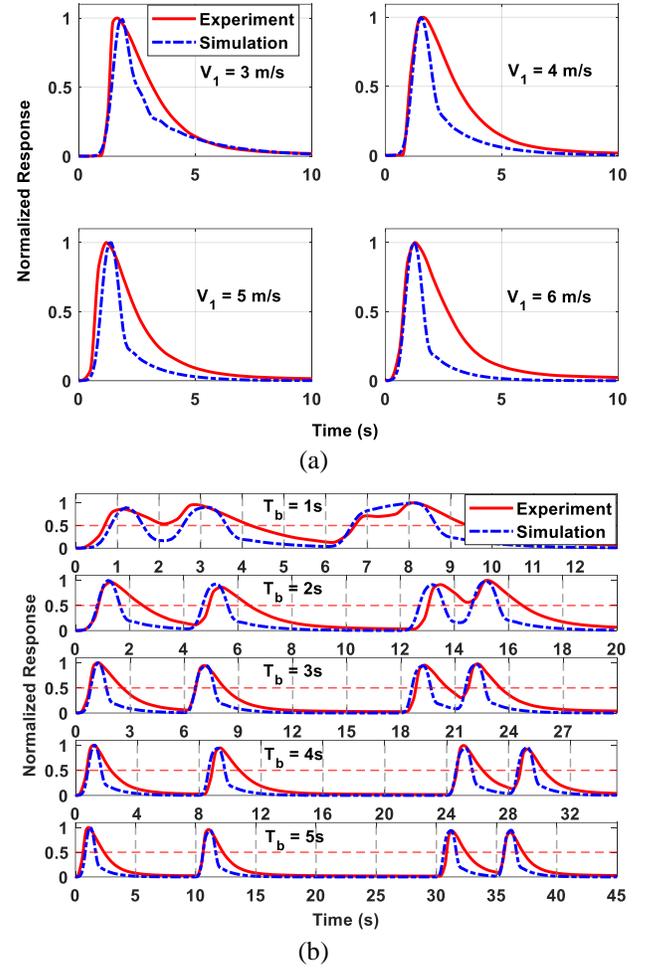

**Fig. 3.** The normalized temporal response using $T_e$=1s and $V_2$= 1m/s for (a) various airflow velocities at inlet 1 (V₁) after sending one bit '1' and (b) for various bit duration (T_b) after sending one byte '10100011' with V₁ = 6m/s.

In this experiment, we transmitted various binary sequences using different bit durations (T_b) and captured the received signals to decode and reconstruct the transmitted bits. As an example, Fig. 3(b) illustrates the measured and theoretical normalized responses for a transmitted byte (10100011) using different bit durations. The results show strong agreement between experimental and numerical results, considering that the peak time and peak amplitude are the most important parameters for the decoding process. While longer bit durations reduce the bit error rate (BER) and simplify data reconstruction, they also lower the achievable bit rate, necessitating a trade-off between BER and bit rate. To reconstruct transmitted bits, a simple decoder with a fixed threshold can be used at optimized sampling time instants. As demonstrated in Fig. 3(a), the peak time represents the optimal sampling point. Using an airflow speed of 6 m/s as an example, the received signal is decoded under different bit durations. We decode signals by sampling each bit at its respective peak time. Under airflow speed of 6m/s, the peak time of the molecular received signal occurs approximately at 1.25s as can be observed in Fig. 3. Therefore, the sampling times of a binary sequence can be expressed as $T_s = 1.25s + nT_b$ where n=0,1,2...,N is the bit index and N is the total number of the transmitted bits.

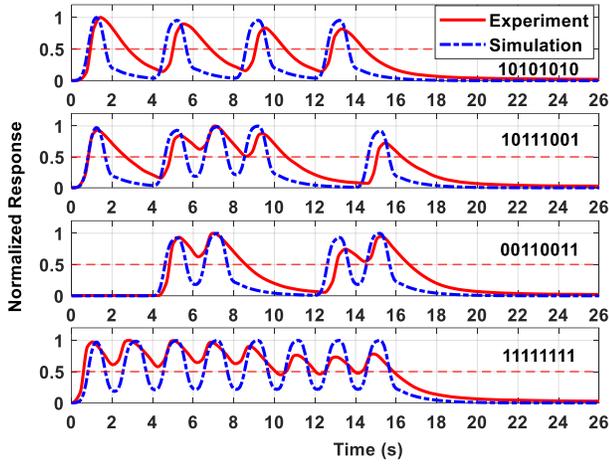

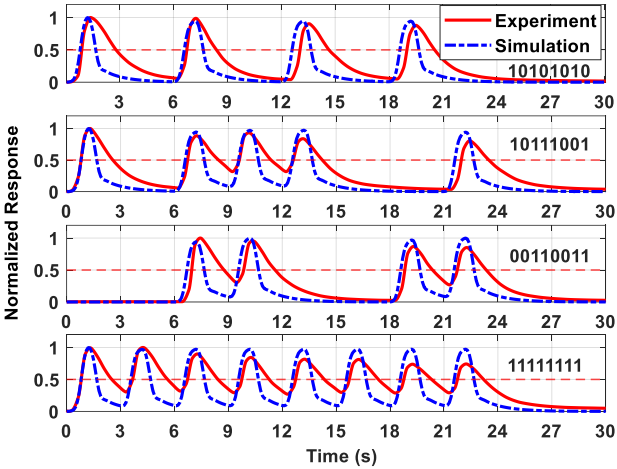

**Fig. 4.** The normalized temporal response after sending various binary sequences using $T_e$=1s, $V_1$= 6m/s, and $V_2$= 1 m/s for (a) bit duration $T_b$=2s and (b) bit duration $T_b$=3s.

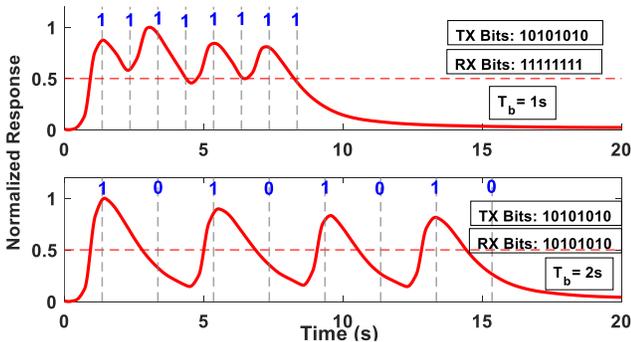

**Fig. 5.** The impact of ISI on decoding the binary sequence (10101010) using the bit duration: $T_b$=1s and $T_b$=2s.

For illustration, Fig. 4 shows a sample of different binary sequences that are transmitted and decoded using various bit durations. To evaluate the performance in terms of BER, a total of 1,024 randomly generated bits (128 bytes) are continuously transmitted and decoded using a threshold value of 0.5 with sampling time $T_s$=1.25s+$nT_b$. Here, a received signal with magnitude $\geq$ 0.5 is decoded as bit 1 while a signal with magnitude < 0.5 is decoded as bit 0. Then the BER is calculated by comparing the decoded bits with the original transmitted bits to find the number of error bits divided by the total sent bits. To investigate the impact of inter-symbol interference (ISI) on decoding the received binary sequence, we show the decoding results for a sample of bits using bit duration of 1s and 2s as shown in Fig. 5. We can observe that the previously transmitted bit will cause interference to the next transmitted bit. This is because a part of the released VOC will continue to reach the receiver in the pipe before releasing the VOC of the next bits. For example, using a bit duration of 1s as shown in Fig. 5a, the transmitted bits 0 are incorrectly decoded as 1's at the sample times. Increasing the bit duration to 2s as shown in Fig. 5b, leads to correctly decoding the transmitted sequence because this gives enough duration for the previously released VOC to be removed from the pipe which reduces the ISI effect. This leads to higher BER for the shorter bit duration due to the ISI that causes errors in decoding the signal at the sampling time. For instance, this system provides a BER of 0.01 and 0.125 using $T_b$=2s and $T_b$=1s, respectively. Therefore, to optimize system performance, channel memory effects and adaptive thresholding techniques must be considered. The system performance in terms of BER and bit rate can be improved using faster airflow velocity, a fast response/recovery sensor, and shorter emission duration.

The experiment was conducted outdoors over several days, during which we observed variations in weather conditions, including humidity and wind fluctuation on the same day. These environmental changes could influence the gas sensor readings, as the sensor is sensitive to temperature and humidity. This may explain the unusual behaviours for some pulses, such as a reduction in peak amplitude and shift in peak time, as illustrated in Figs. 4-5. Additionally, prolonged exposure to airflow and VOC might alter the sensor's response over time. Despite these factors, the system maintains satisfactory performance, even when transmitting a long sequence of bits continuously.

We have successfully validated the developed theoretical model using the experimental testbed. Next, we will conduct a more in-depth analysis of the proposed MC system to further explore its performance and behaviour. First, we will analyze how the emission duration of the VOC signal influences the molecular signal received by the sensor. The temporal molecular received signal is plotted for different emission durations of the information signal (i.e., the VOC pulses) at the TX while keeping all other parameters constant, as shown in Fig. 6. The received molecular signal rises over time until reaching its peak amplitude, after which it gradually decays to a negligible level. The observed rise and decay pattern aligns with theoretical expectations for such molecular communication channels. This happens because the amount of VOC that reaches the sensor continues to increase over time, and after the emission stops, it begins to decrease. We observe that the peak amplitude of the received molecular signal increases with longer emission durations. This is because a longer emission period results in a larger VOC amount being released into the pipe, resulting in a higher concentration reaching the VOC sensor at RX. For example, the peak amplitude shows an increase approximately by a factor of 1.6, 2, and 2.25 when the emission duration is increased from 0.5s to 1s, 1.5s, and 2s, respectively. This demonstrates that longer emission times lead to proportionally higher signal strengths at RX. Furthermore, the molecular signal initially shows zero value because the VOC signal requires time to travel through the pipe before reaching the RX, which is positioned at a considerable distance from TX.

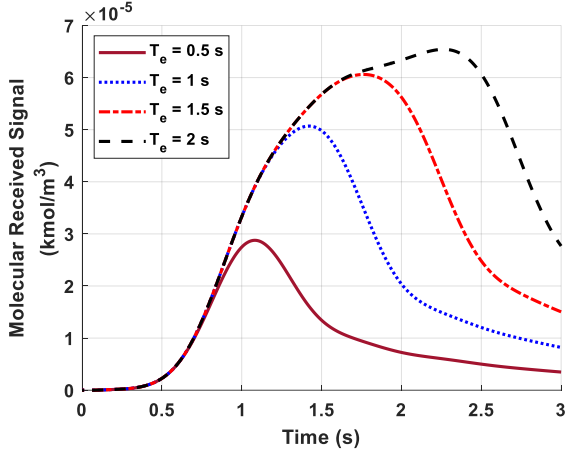

**Fig. 6.** Molecular received signal for various emission durations ($T_e$) of VOC signal using $V_1 = 5$ m/s and $V_2 = 2.5$ m/s.

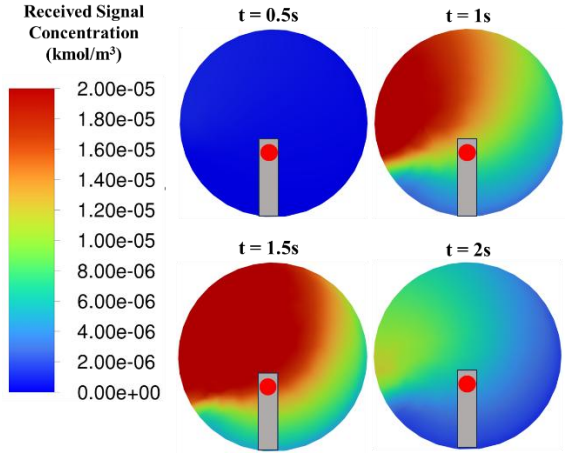

**Fig. 7.** The colourmap of VOC concentration contour on the cross-sectional YZ-plane in the vicinity of the RX at different time instances using the following parameters: $V_1 = 5$m/s, $V_2 = 5$m/s, and $T_e = 1$s.

Extending the emission period increases the peak time of the received signal. However, further increases in emission duration do not significantly raise the peak amplitude but instead result in a flatter amplitude pattern. This happens because the newly arrived VOC molecules at RX will extend the signal amplitude but at the same time the bulk airflow will clean the previously arrived gas molecules. This also can be observed from the concentration contour in the vicinity of the RX at different time instances as shown in Fig. 7. This is an important fact for MC systems that use amplitude-based detection techniques because the decoding process depends mainly on the amplitude characteristic of the received signal. Using a longer emission duration allows VOC sensors sufficient time to detect molecular signals, specifically beneficial for sensors with low sensitivity and slow response times. However, long emission duration can lead to ISI, necessitating a larger bit duration ($T_b$) to compensate. This increases the BER and reduces the overall bit rate.

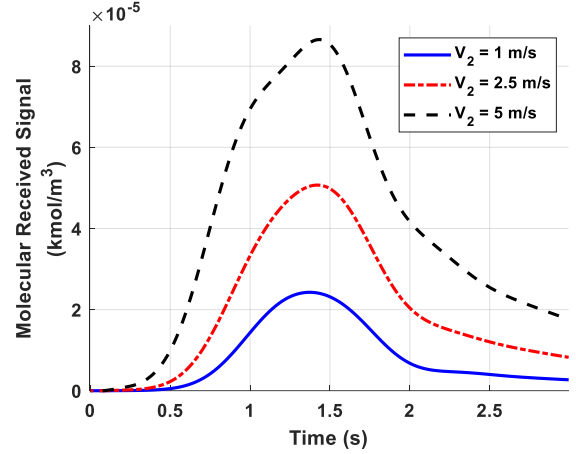

**Fig. 8.** Molecular received signal for various velocities of VOC emission at inlet-2 using $V_1 = 5$ m/s and $T_e = 1$s.

We also studied how the VOC flow velocity, induced by the sprayer in the branch pipe (inlet 2), affects the molecular received signal, as shown in Fig. 8. The results show that the VOC flow velocity significantly affects the peak amplitude of the received signal. Increasing the flow velocity ($V_2$) in the branch pipe increases the peak amplitude of the received signal. For example, when the VOC flow velocity increases from 1 m/s to 2.5 m/s, the peak amplitude doubles, while a further increase to 5 m/s results in a 3.5-fold enhancement in the peak amplitude. The observed increase in signal amplitude can be attributed to two key factors. First, higher emission velocities lead to greater VOC release into the main pipe, increasing mass/molar flow rates (as described by Eqs. 12 and 13) and consequently higher VOC concentrations at the RX. Second, higher emission velocities drive the VOC gas toward the pipe center. This enhances gas mixing within the main pipe and produces a higher dilution factor (DF), as quantified by Eq. 15. Therefore, this significantly increases the likelihood of VOC molecules reaching RX since the gas sensor is positioned at the pipe center, as shown in Fig. 2(a).

$$DF = (V_2/V_1) \times (D_2/D_1)^2 \qquad (15)$$

where $V_1$ is the airflow velocity at inlet 1, $V_2$ is the VOC flow velocity at inlet 2, $D_1$ is the diameter of the main pipe, and $D_2$ is the diameter of the branch pipe.

Figure 9 shows the VOC concentration contour for different VOC flow velocities ($V_2$) in both pipe segments illustrated in Fig. 2(a). The YZ-plane corresponds to a vertical slice from the first pipe segment, while the YX-plane represents a vertical slice from the second pipe segment. When the VOC flow velocity in the branch pipe decreases (i.e., lower DF), the released VOC gas becomes more easily deflected by the carrier airflow in the main pipe, as shown in Fig. 9. As a result, the VOC gas remains confined near the upper part of the pipe wall, reducing the amount detected by the gas sensor. However, the VOC flow velocity ($V_2$) has a negligible effect on the peak time. This is because the VOC flow is applied for a short duration determined by the emission duration. Thus, it does not significantly impact the gas mixture velocity in the main pipe ($V_1$). In other words, the gas flow in the main pipe is mainly determined by the airflow at inlet 1. It is worth mentioning that the location of the TX and the RX with respect to the main pipe may affect the received signal. For example, we can see that the VOC concentration is not uniformly distributed in the cross-section of the pipe as illustrated in Fig. 7. Therefore, the sensor must be placed at a sufficient vertical height to be in an area of high VOC concentration. However, this also depends on the orientation of the branch pipe which will also lead to different concentration contours in the main pipe. In this work, we assume vertical orientation for the branch pipe with respect to the main pipe for the sake of analysis.

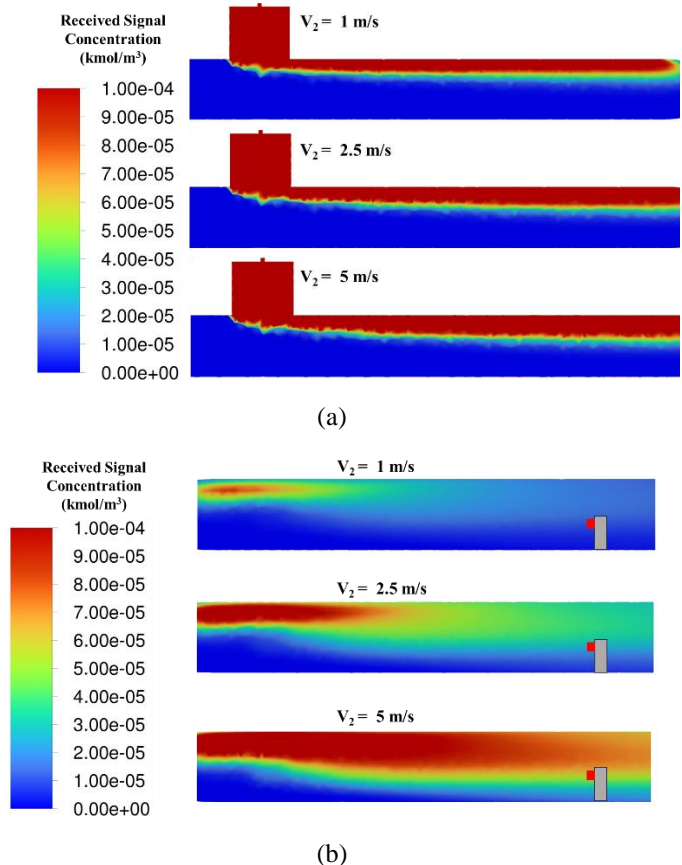

**(a)**

**(b)**

**Fig. 9.** The colourmap of VOC concentration contour for different VOC flow velocities ($V_2$) with $V_1$ = 5m/s and $T_e$ = 1s at time instant t = 1s on (a) the vertical YZ-plane of the pipe (b) on the vertical YX-plane of the pipe.

The impact of the velocity of the carrier signal (airflow) in the main pipe on the molecular received signal is shown in Fig. 10. In the first few milliseconds, the molecular received signal increases slowly because the information signal (VOC) takes time to reach RX, especially at lower airflow velocities. The results demonstrate that airflow velocity significantly affects both the magnitude and timing characteristics of the received signal. The peak amplitude decreases as the airflow velocity increases and vice versa. This happens because as the airflow velocity increases (i.e., the DF decreases), the airflow will drift the VOC emitted from the sprayer faster. Thus, the VOC will move close to the upper side of the main pipe wall, reducing the amount of VOC gas that reaches the gas sensor. On the other hand, increasing the airflow velocity will shorten the peak time of the received signal since the VOC reaches the RX faster. For example, as the airflow velocity in the main pipe rises from 1 m/s to 2.5 m/s and then to 5 m/s, the peak amplitude drops by roughly half and one-fourth, respectively, while the peak time shortens by approximately 1.8 times and 2 times, respectively. Using high airflow velocity will speed up receiving the VOC signal by the sensor which in order allows us to use a shorter bit duration and thus achieve a higher bit rate. However, the higher airflow velocity will make the sensing process by the VOC sensor harder. This is because the VOC signal reaching the sensor will be weak and fast which requires a sensor with high sensitivity and short response time.

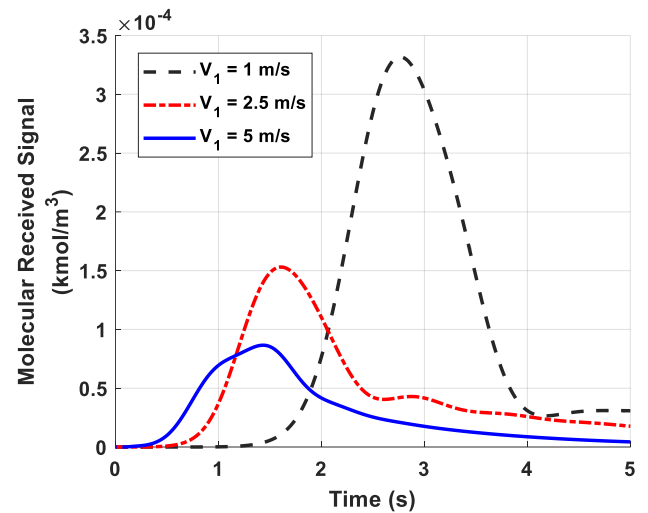

**Fig. 10.** Molecular received signal using $V_2$ = 5m/s and $T_e$ = 1s for different airflow velocities at inlet 1 ($V_1$).

## VI. CONCLUSIONS

In this work, we propose a macroscale MC system for IoT-based in-pipe inspection and monitoring networks to establish communication links between sensor nodes and inspection robots inside pipelines. The system can be used to provide communication in three key scenarios: between sensor nodes, between a control/monitoring unit and an in-pipe robot, and between a sensor node and an in-pipe robot. To analyze the system, we develop a mathematical model using the CFD approach, accounting for various parameters that influence the molecular received signal at RX. Additionally, we implement an experimental testbed to validate the theoretical model and assess the system's feasibility by transmitting and reconstructing binary sequences using VOC gas as the information signal. The irregular flow within the pipe is modelled as a transient 3D turbulent flow to replicate real-world conditions. Our study investigates the impact of the following key parameters on the received signal: emission duration, information signal velocity, carrier signal velocity, and bit duration. The results demonstrate that these parameters significantly affect the temporal and magnitude characteristics of the received signal, highlighting the need for optimal parameter selection to achieve optimal performance. This work serves as a preliminary step toward advancing macroscale MC applications to address communication challenges in pipeline-based IoT monitoring and inspection networks. The proposed system is also adaptable to other molecular signals and modulation techniques with minimal adjustments.

## REFERENCES

[1] S. Folga, *Natural Gas Pipeline Technology Overview*. Argonne, IL, USA: Argonne National Laboratory, 2007.

[2] A. Verma, A. Kaiwart, N. D. Dubey, F. Naseer, and S. Pradhan, "A review on various types of in-pipe inspection robot," *Materials Today: Proceedings*, vol. 50, 5, pp. 1425-1434, 2022.

[3] J. Liang, J. Tu, and V. C. Leung, "Mobile sensor deployment optimization algorithm for maximizing monitoring capacity of large-scale acyclic directed pipeline networks in smart cities," *IEEE Internet of Things Journal*, vol. 8, no. 21, pp. 16083-16095, 2020.

[4] A. Albaseer, U. Baroudi, and S. Selim, "Recursive Clustering Approach for Wireless Sensor Networks in Pipeline Monitoring Application," *IEEE Sensors Journal*, 2024.

[5] M. R. Islam, S. Azam, B. Shanmugam, and D. Mathur, "An Intelligent IoT and ML-Based Water Leakage Detection System," *IEEE Access*, vol. 11, pp. 123625 - 123649, 2023.

[6] F. Karray, A. Garcia-Ortiz, M. W. Jmal, A. M. Obeid, and M. Abid, "Earnpipe: A testbed for smart water pipeline monitoring using wireless sensor network," *Procedia Computer Science*, vol. 96, pp. 285-294, 2016.

[7] K. Shaheen, A. Chawla, F. E. Uilhoorn, and P. S. Rossi, "Partial-Distributed Architecture for Multi-Sensor Fault Detection, Isolation and Accommodation in Hydrogen-Blended Natural Gas Pipelines," *IEEE Internet of Things Journal*, 2024.

[8] J. Liang, H. Zhang, X. Deng, and Z. He, "On zone-differentiated time-constrained flow capacity intelligent monitoring for large-scale urban pipeline systems by mobile sensors," *IEEE Internet of Things Journal*, vol. 9, no. 23, pp. 23599-23613, 2022.

[9] C. Spandonidis, P. Theodoropoulos, and F. Giannopoulos, "A combined semi-supervised deep learning method for oil leak detection in pipelines using IIoT at the edge," *Sensors*, vol. 22, no. 11, p. 4105, 2022.

[10] J. Wan, Y. Yu, Y. Wu, R. Feng, and N. Yu, "Hierarchical leak detection and localization method in natural gas pipeline monitoring sensor networks," *Sensors*, vol. 12, no. 1, pp. 189-214, 2011.

[11] J. Korsimaa *et al.*, "Wireless and battery-operatable IoT platform for cost-effective detection of fouling in industrial equipment," *Scientific Reports*, vol. 14, no. 1, p. 14084, 2024.

[12] Beilai Technology. "Cellular IoT M2M RTU." bliiot.com. https://www.bliiot.com/m2m-rtu-p00175p1.html (accessed Sep. 1, 2024).

[13] A. R. Silva and M. Moghaddam, "Design and implementation of low-power and mid-range magnetic-induction-based wireless underground sensor networks," *IEEE Transactions on Instrumentation and Measurement*, vol. 65, no. 4, pp. 821-835, 2015.

[14] N. Chaamwe, W. Liu, and H. Jiang, "Wave propagation communication models for wireless underground sensor networks," in *2010 IEEE 12th international conference on communication technology*, 2010, pp. 9-12.

[15] S. Mekid, D. Wu, R. Hussain, and K. Youcef-Toumi, "Channel modeling and testing of wireless transmission for underground in-pipe leak and material loss detection," *International Journal of Distributed Sensor Networks*, vol. 13, no. 11, pp. 1-16, 2017.

[16] T. Nagashima, Y. Tanaka, and S. Ishihara, "Measurement of wireless communication characteristics in sewer pipes for sewer inspection systems using multiple wireless sensor nodes," in *2015 IEEE 26th Annual International Symposium on Personal, Indoor, and Mobile Radio Communications (PIMRC)*, 2015, pp. 2055-2060.

[17] W. Zhao *et al.*, "A coordinated wheeled gas pipeline robot chain system based on visible light relay communication and illuminance assessment," *Sensors*, vol. 19, no. 10, p. 2322, 2019.

[18] W. Zhao *et al.*, "A preliminary experimental study on control technology of pipeline robots based on visible light communication," in *2019 IEEE/SICE International Symposium on System Integration (SII)*, 2019, pp. 22-27.

[19] M. M. Al-Zu'bi and A. M. Sanagavarapu, "Modeling a composite molecular communication channel," *IEEE Transactions on Communications*, vol. 66, no. 8, pp. 3420-3433, 2018.

[20] M. M. Al-Zu'bi and A. S. Mohan, "Modelling of implantable drug delivery system in tumor microenvironment using molecular communication paradigm," *IEEE Access*, vol. 7, pp. 141929-141940, 2019.

[21] Y. Cevallos *et al.*, *Molecular Communications: An Analysis from Networking Theories Perspective*. Springer Nature, 2023.

[22] T. Nakano, A. W. Eckford, and T. Haraguchi, *Molecular Communication*. Cambridge, United Kingdom: Cambridge University Press, 2013.

[23] M. M. Al-Zubi, A. S. Mohan, P. Plapper, and S. H. Ling, "Intrabody molecular communication via blood-tissue barrier for internet of bio-nano things," *IEEE Internet of Things Journal*, vol. 9, no. 21, pp. 21802-21810, 2022.

[24] L. Felicetti, M. Femminella, G. Reali, and P. Liò, "Applications of molecular communications to medicine: A survey," *Nano Communication Networks*, vol. 7, pp. 27-45, 2016.

[25] F. Dinc, B. C. Akdeniz, A. E. Pusane, and T. Tugcu, "A general analytical approximation to impulse response of 3-D microfluidic channels in molecular communication," *IEEE transactions on nanobioscience*, vol. 18, no. 3, pp. 396-403, 2019.

[26] S. Giannoukos, D. T. McGuiness, A. Marshall, J. Smith, and S. Taylor, "A chemical alphabet for macromolecular communications," *Analytical chemistry*, vol. 90, no. 12, pp. 7739-7746, 2018.

[27] S. Giannoukos, A. Marshall, S. Taylor, and J. Smith, "Molecular communication over gas stream channels using portable mass spectrometry," *Journal of The American Society for Mass Spectrometry*, vol. 28, no. 11, pp. 2371-2383, 2017.

[28] H. Unterweger *et al.*, "Experimental molecular communication testbed based on magnetic nanoparticles in duct flow," in *2018 IEEE 19th International Workshop on Signal Processing Advances in Wireless Communications (SPAWC)*, 2018, pp. 1-5.

[29] S. Wang, W. Guo, S. Qiu, and M. D. McDonnell, "Performance of macro-scale molecular communications with sensor cleanse time," in *2014 21st International Conference on Telecommunications (ICT)*, 2014, pp. 363-368.

[30] W. Guo *et al.*, "Molecular communications: Channel model and physical layer techniques," *IEEE Wireless Communications*, vol. 23, no. 4, pp. 120-127, 2016.

[31] N. Farsad, W. Guo, and A. W. Eckford, "Tabletop molecular communication: Text messages through chemical signals," *PloS one*, vol. 8, no. 12, p. e82935, 2013.

[32] D. T. McGuiness, S. Giannoukos, A. Marshall, and S. Taylor, "Parameter analysis in macro-scale molecular communications using advection-diffusion," *IEEE Access*, vol. 6, pp. 46706-46717, 2018.

[33] D. T. Mcguiness, S. Giannoukos, S. Taylor, and A. Marshall, "Experimental and analytical analysis of macro-scale molecular communications within closed boundaries," *IEEE Transactions on Molecular, Biological and Multi-Scale Communications*, vol. 5, no. 1, pp. 44-55, 2019.

[34] L. Khaloopour *et al.*, "An experimental platform for macro-scale fluidic medium molecular communication," *IEEE Transactions on Molecular, Biological and Multi-Scale Communications*, vol. 5, no. 3, pp. 163-175, 2019.

[35] Y. Huang, M. Wen, L.-L. Yang, C.-B. Chae, X. Chen, and Y. Tang, "Space shift keying for molecular communication: Theory and experiment," in *2019 IEEE Global Communications Conference (GLOBECOM)*, 2019, pp. 1-6.

[36] S. Bhattacharjee *et al.*, "A testbed and simulation framework for air-based molecular communication using fluorescein," in *Proceedings of the 7th ACM International Conference on Nanoscale Computing and Communication*, 2020, pp. 1-6.

[37] M. Abbaszadeh *et al.*, "Mutual information and noise distributions of molecular signals using laser induced fluorescence," in *2019 IEEE Global Communications Conference (GLOBECOM)*, 2019, pp. 1-6.

[38] I. Atthanayake, S. Esfahani, P. Denissenko, I. Guymer, P. J. Thomas, and W. Guo, "Experimental molecular communications in obstacle rich fluids," in *Proceedings of the 5th ACM International Conference on Nanoscale Computing and Communication*, 2018, pp. 1-2.

[39] M. Ozmen, E. Kennedy, J. Rose, P. Shakya, J. K. Rosenstein, and C. Rose, "High speed chemical vapor communication using photoionization detectors in turbulent flow," *IEEE Transactions on Molecular, Biological and Multi-Scale Communications,* vol. 4, no. 3, pp. 160-170, 2018.

[40] Ansys Inc., "Ansys Fluent Theory Guide," Ansys Inc., Canonsburg, PA, USA, 2021.

[41] B. E. Launder and D. B. Spalding, *Lectures in Mathematical Models of Turbulence*. London, United Kingdom: Academic Press, 1972.

[42] B. E. Launder and D. B. Spalding, "The numerical computation of turbulent flows," *Computer Methods in Applied Mechanics and Engineering,* vol. 3, no. 2, pp. 269-289, 1974.

[43] S. C. Johnson & Son Inc. "Glade® Automatic Spray Refill - Peony and Berry Bliss." whatsinsidescjohnson.com. https://www.whatsinsidescjohnson.com/ph/en/brands/glade/Glade--Automatic-Spray-Refill---Peony-and-Berry-Bliss (accessed 1 sep., 2024).

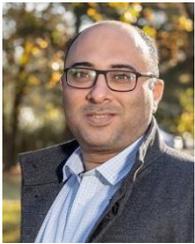

**Muneer M. Al-ZuBi** (Member, IEEE) received his Ph.D. degree in engineering from the University of Technology Sydney (UTS), Sydney, Australia, in 2020. He worked as a Research Associate with the Center of Excellence for Innovative Projects, Jordan University of Science and Technology (JUST), from 2020 to 2021. From 2021 to 2022, he was a postdoctoral researcher with the Department of Engineering, University of Luxembourg, Luxembourg, and also, he was a remote visiting scholar with the School of Electrical and Data Engineering, UTS. He is currently a Postdoctoral Researcher at the Communication Theory Lab (CTL), King Abdullah University of Science and Technology, Saudi Arabia. His research interests lie in the areas of wireless communication, EM wave propagation, and molecular communication.

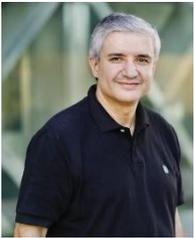

**Mohamed-Slim Alouini** (Fellow, IEEE) received his Ph.D. degree in electrical engineering from the California Institute of Technology (Caltech), Pasadena, CA, USA, in 1998. He served as a faculty member at the University of Minnesota, Minneapolis, MN, USA, then at Texas A&M University at Qatar, Education City, Doha, Qatar before joining King Abdullah University of Science and Technology (KAUST), Thuwal, Makkah Province, Saudi Arabia as a professor of electrical engineering in 2009. His current research interests include the modeling, design, and performance analysis of wireless communication systems.